\documentclass[12pt]{article}
\usepackage{amsmath}
\usepackage{amssymb}
\usepackage[capitalize]{cleveref}
\usepackage{mathrsfs} 
\usepackage[left=2.5cm,top=2.50cm,right=2.5cm,bottom=2.50cm]{geometry}
\usepackage{amsmath,amssymb,latexsym,color,cancel,graphicx,bbm,colortbl}
\usepackage[english]{babel}
\usepackage[utf8]{inputenc}
\usepackage{ragged2e}
\usepackage{cite}
\usepackage{placeins}
\usepackage{float}
\begin{document}
\date{}

\title{An algebraic solution of Dirac equation on a static curved space-time}
\author{M. Salazar-Ram\'irez$^{a}$\footnote{{\it E-mail address:} msalazarra@ipn.mx},  R. D. Mota$^{b}$, D. Ojeda-Guill\'en$^{a}$, A. Gonz\'alez-Cisneros$^{a}$\\ } \maketitle

\begin{minipage}{0.9\textwidth}
\small $^{a}$ Escuela Superior de C\'omputo, Instituto Polit\'ecnico Nacional,
Av. Juan de Dios B\'atiz esq. Av. Miguel Oth\'on de Mendiz\'abal, Col. Lindavista,
Alc. Gustavo A. Madero, C.P. 07738, Ciudad de M\'exico, Mexico.\\
\small $^{b}$ Escuela Superior de Ingenier{\'i}a Mec\'anica y El\'ectrica, Unidad Culhuac\'an,
Instituto Polit\'ecnico Nacional, Av. Santa Ana No. 1000, Col. San Francisco Culhuac\'an, Alc.  Coyoac\'an, C.P. 04430, Ciudad de M\'exico, Mexico.\\
\end{minipage}

\begin{abstract}
We present exact solutions of the Dirac equation in static curved space-time using two distinct algebraic approaches. The first method employs $su(1,1)$ algebra operators together with the tilting transformation, enabling the derivation of the energy spectrum and eigenfunctions for both the Hydrogen atom and the Dirac-Morse oscillator. The second approach, based on the Schr\"odinger factorization method, extends the analysis to three representative potentials: the hydrogen atom, the Dirac-Morse oscillator, and a linear radial potential. Although structurally different from those obtained in the first method, the resulting operators in this approach also close the $su(1,1)$ algebra and, through representation theory, yield the corresponding energy spectra and eigenfunctions.
\end{abstract}


\section{Introduction}

The Dirac equation, a fundamental framework of relativistic quantum mechanics, has been extensively investigated using a variety of analytical techniques, including the factorization method, the Nikiforov-Uvarov formalism, shape invariance methods, and approaches grounded in supersymmetric quantum mechanics~\cite{Mon,Mio1,Hao,Zar,Ikot}. These methodologies have enabled its application to a broad spectrum of interaction potentials, such as the generalized Kratzer, Morse, Cornell, generalized P\"oschl--Teller and Mie-type potentials\cite{DeOliv4,Alex,Abu,Jia,Aydo}.

Since Parker's pioneering analysis of a one-electron atom in an arbitrary curved space-time~\cite{Par}, wherein he extended the Dirac equation to curved geometries and developed a perturbation theory for degenerate stationary states under the Hermiticity conditions dictated by the space-time curvature, there has been a growing interest in the quantum dynamics of such systems\cite{Fig,Hos, Bir,Aud, Aud1, Par1, Par2, Par3}.

In particular, by expressing the Dirac Hamiltonian in Fermi normal coordinates and expanding it to first order in the Riemann curvature tensor---including corrections to the electromagnetic interaction---Parker laid the groundwork for subsequent studies of quantum fields in curved backgrounds.

Reference \cite{Yesi} establishes a connection between Dirac Hamiltonians formulated within the frameworks of minimal length and gravitational effects, allowing for the derivation of exact solutions in both scenarios. By employing a factorization approach, the associated partner potentials are constructed. Furthermore, transformations linking Dirac systems in flat and curved space-time naturally give rise to a metric-dependent shape function, demonstrating that point transformations can be effectively utilized to derive energy-dependent potentials.

De Oliveira, M. D, explores the spherically symmetric Dirac equation in curved space-time with vector potentials $V(r)$ and tensor $A(r)$ under the influence of a scalar potential $U(r)$. Using a unitary transformation, we decouple the radial components, obtaining an effective Schr\"odinger equation in the framework of supersymmetric quantum mechanics (MQ SUSY). For $U(r)=0$, the potentials can be expressed by a superpotential $W(r)$, which allows us to obtain exact solutions for potentials of invariant form in curved space-time \cite{DEOLIV2}.

Recent research extends the investigation of the relativistic Dirac--Rosen--Morse problem by incorporating curved space-time effects, described by the metric $ds^2 = \left(1 + \alpha^2 U(r)\right)^2 (dt^2 - dr^2) - r^2 d\theta^2 - r^2 \sin^2\theta\, d\phi^2$, where \( \alpha \) is the fine-structure constant and \( U(r) \) is a scalar potential. The system includes an electromagnetic field \( A_\mu = \left(V(r),\, cA(r),\, 0,\, 0\right) \), and spherical symmetry allows the angular spinor components to be expressed in terms of spherical harmonics. A unitary transformation of the radial spinor enables the vector potential \( A(r) \) to be written as a function of \( V(r) \) and \( U(r) \), facilitating analytical solutions to the Dirac--Rosen--Morse system in curved space-time.

Furthermore, the study presents a graphical comparison of eigenenergies and probability densities in curved versus flat space-time, highlighting how space-time curvature modifies the two-component radial spinor. The upper and lower components correspond to particle and antiparticle states, respectively~\cite{deoliv3}.

M.D. de Oliveira and Alexandre G.M. Schmidt base their investigation on the Dirac equation in static curved space-time, employing a static spherically symmetric metric characterized by the functions \( f(r) \) and \( g(r) \). By decoupling the radial and angular components of the Dirac equation, the authors derive exact analytical solutions for a variety of quantum systems, including the hydrogen atom, the Dirac--Morse oscillator, and a linearly confining radial potential. A key result demonstrates how spinor solutions in curved space-time are connected to their flat-space-time counterparts through the temporal metric component \( g_{tt} \)~\cite{GMS}.

Within an algebraic framework in Ref \cite{Mio1}, the modified Dirac oscillator in curved space-time is studied, with special attention to its spin and pseudo-spin symmetry limits. By systematically applying the Schr\"odinger factorization method to the decoupled second-order radial equations, the study explicitly constructs the Lie algebra generators of $su(1,1)$. The analysis provides complete energy spectra for both symmetry regimes, demonstrating how the algebraic structure determines the quantum mechanical solutions.

The structure of this paper is as follows: In Section 2, we begin with a brief review of the essential formalism underlying the Dirac equation in curved space-time. We then extend this framework to include an external electromagnetic field $A_\mu$ and focus on the corresponding radial equation. To obtain exact eigenfunctions, we impose the condition $e^{f(r)} = e^{g(r)} = 1 + \alpha^2 U(r)$, where $U(r)$ is a well-behaved function with radial dependence, and $\alpha$ denotes the fine-structure constant. Subsequently, the radial components of the spinor are decoupled, leading to radial wave functions that depend explicitly on both the choice of $U(r)$ and the electromagnetic field $A_\mu$.

In Section 3, by introducing a set of operators that span the Lie algebra $su(1,1)$ along with the tilting transformation, we derive the energy spectrum and eigenfunctions of the hydrogen atom in curved space-time from an algebraic perspective. In Section 4, we explore the Dirac Morse oscillator in curved space-time through an algebraic framework, where the implementation of a tilting transformation and a set of operators forming the $su(1,1)$ Lie algebra enables the determination of both the energy spectrum and the corresponding eigenfunctions.

In sections $5$, $6$ and $7$ the radial Hamiltonians for the linear radial potential, the Dirac Morse oscillator, and the hydrogen atom in curved space-time are treated using the Schr\"odinger factorization technique, which enables the construction of the three generators associated with the $su(1,1)$ Lie algebra. The energy spectrum is calculated within the framework of the theory of irreducible representations. Additionally, an analytical solution for the radial wave functions is obtained. Finally, in Sec. $8$, we present some conclusions regarding our work.

\section{A Review of the Dirac Equation in Curved space-time in spherical coordinates on the presence of electromagnetic field}
We begin our analysis by considering a spherically symmetric metric described by the line element
\begin{equation}
ds^2 = e^{2f(r)} dt^2 - e^{2g(r)} dr^2 - r^2 d\theta^2 - r^2 \sin^2{\theta} d\phi^2,
\end{equation}
where \( f(r) \) and \( g(r) \) are arbitrary functions of the radial coordinate. Under the gravitational potential ansatz \( e^f = e^g = 1 + U(r) \), the space-time metric takes the general relativistic form
\begin{align}\label{eline}
ds^2 = \left(1 + \frac{U(r)}{c^2} \right)^2 (dt^2 - dr^2) - r^2 d\theta^2 - r^2 \sin^2 \theta\, d\phi^2,
\end{align}
where \( U(r) \) is a well-behaved scalar potential, and \( \alpha \) denotes the fine-structure constant. Upon decoupling the radial component of the Dirac spinor, the resulting radial wave functions are found to depend explicitly on both the form of \( U(r) \) and the electromagnetic four-potential \( A_\mu \), which encapsulates the scalar and vector interactions governing the space-time curvature.
For an electromagnetic field described by the four-potential \( A^\mu = \big(V(r), cA(r), 0, 0\big) \), the Dirac eigenvalue equation \( H_d \Psi_d = E \Psi_d \) yields the Hamiltonian
\begin{align}\label{ham1}
H_d &= -i c\, e^{f - g} \alpha^1 \left( \partial_r + \frac{1}{r} + \frac{f'_r}{2} + i A(r) \right) \nonumber\\
&\quad - i c\, \frac{e^f}{r} \left\{ \alpha^2_d \left( \partial_\theta + \frac{\cot \theta}{2} \right) + \frac{\alpha^3}{\sin \theta} \partial_\phi \right\} + c^2 \beta e^f + V(r),
\end{align}
where \( \alpha^i \) and \( \beta \) are the standard Dirac matrices. To simplify the analysis of our system, we now introduce a similarity transformation. If \( H' = U (H_d / c) U^{\dagger} \) and \( \psi' = U \psi_d \), then the angular component operator is defined as
\begin{equation}
\hat{S} = \sigma^1 \left( \partial_\theta + \frac{\cot\theta}{2} \right) + \sigma^2 \frac{\partial_\phi}{\sin\theta},
\end{equation}
where the matrices \( \alpha^i \) take the standard representation
\begin{equation}
\alpha^i =
\begin{pmatrix}
0 & \sigma^i \\
\sigma^i & 0
\end{pmatrix}.
\end{equation}
We adopt the spinor ansatz
\begin{equation}
\Psi'(r, \theta, \phi) =
\begin{pmatrix}
R_1(r) \chi(\theta, \phi) \\
-i \sigma^3 R_2(r) \chi(\theta, \phi)
\end{pmatrix},
\end{equation}
where \( R_1(r) \) and \( R_2(r) \) denote the radial components and \( \chi(\theta, \phi) \) represents the angular spinor. This leads to the following expression
\begin{align}\nonumber
&\begin{pmatrix} e^f + \alpha^2 V(r) - \epsilon & -i\alpha e^{f-g} \sigma^3 \left[ \partial_r + \frac{1}{r} + \frac{f'_r}{2} - A_r \right] - i\alpha \frac{e^f}{r}\hat{S} \\ -i\alpha e^{f-g} \sigma^3 \left[ \partial_r + \frac{1}{r} + \frac{f'_r}{2} + A_r \right] - i\alpha \frac{e^f}{r} \hat{S} & -e^f + \alpha^2 V(r) - \epsilon \end{pmatrix}\times\\ &\begin{pmatrix} R_1(r) \chi(\theta,\phi) \\ -i\sigma^3 R_2(r) \chi(\theta,\phi) \end{pmatrix} = 0.
\end{align}
The electromagnetic coupling term, represented by $\sigma^3 A_r$, appears in the main diagonal of the Hamiltonian. In this formulation, the antiparticle sector assumes the form $+i\sigma^3 A_r$, while the particle sector takes the form $-i\sigma^3 A_r$, leading to the following coupled equations
\begin{align}\label{dif1o}
&\left( \alpha^2 V + m e^{f} - \epsilon \right) R_1 \chi
- \alpha e^{f-g} \left[ \partial_r + \frac{1}{r} + \frac{f'}{2} - A_r \right] R_2 \chi
- i\alpha \frac{e^{f}}{r} R_2 (\hat{S}\sigma^3) \chi = 0, \\\label{dif2o}
&\left( \alpha^2 V - m e^{f} - \epsilon \right) R_2 \chi
+ \alpha e^{f-g} \left[ \partial_r + \frac{1}{r} + \frac{f'}{2} + A_r \right] R_1 \chi
- i\alpha \frac{e^{f}}{r} R_1 (\hat{S}\sigma^3) \chi = 0.
\end{align}
After a detailed analysis, the four-component spinorial wavefunction takes the form
\begin{align}
\Psi_c(r,\theta,\phi) = \mathcal{N}
\begin{pmatrix}
R_1(r) \, \mathcal{Y}^{|m|j}_{j + 1/2}(\theta, \phi) \\
i R_2(r) \, \mathcal{Y}^{|m|j}_{j - 1/2}(\theta, \phi)
\end{pmatrix},
\end{align}
where $\mathcal{Y}^{|m|j}$ denotes the spinor spherical harmonics, defined by
\begin{equation}
\mathcal{Y}^{j = l \pm \frac{1}{2}, m}_l(\theta, \phi) =
\frac{1}{\sqrt{2l + 1}}
\begin{pmatrix}
\pm \sqrt{l \pm m + \frac{1}{2}} \, Y^{m - \frac{1}{2}}_l(\theta, \phi) \\
\sqrt{l \mp m + \frac{1}{2}} \, Y^{m + \frac{1}{2}}_l(\theta, \phi)
\end{pmatrix}.
\end{equation}
In this decomposition, the orbital angular momentum quantum number \( l \) can take two possible values for a given total angular momentum \( j \), \( l = j - \frac{1}{2} \) (aligned case) or \( l = j + \frac{1}{2} \) (anti-aligned case), where \( Y^{m}_l(\theta, \phi) \) are the standard spherical harmonics, and \( \mathcal{N}\) is a normalization constant.
The radial functions \( F_{1}(r) \) and \( F_{2}(r) \), corresponding to the two spinor components, are obtained by solving the coupled radial differential equations derived from the Dirac equation\cite{GMS}, thereby giving the following equation
\begin{equation}\label{hamprinc}
\begin{pmatrix}
1 + \alpha^2\left(V(r)+U(r)\right) & -\alpha\left[\frac{d}{dr} - \frac{\lambda}{r}\left(1 + \alpha U(r)\right) - A_r \right] \\\\
-\alpha\left[\frac{d}{dr} - \frac{\lambda}{r}\left(1 + \alpha U(r)\right) - A_r \right] & -1 + \alpha^2\left(V(r)-U(r)\right)
\end{pmatrix}
\begin{pmatrix}
F_{1}(r) \\\\
F_{2}(r)
\end{pmatrix}
= \epsilon\begin{pmatrix}
F_{1}(r) \\\\
F_{2}(r)
\end{pmatrix},
\end{equation}
where the substitutions \( R_1 = \frac{F_1(r)}{r} e^{-f/2} \) and \( R_2 = \frac{F_2(r)}{r} e^{-f/2} \) have been made in Eqs.(\ref{dif1o}) and (\ref{dif2o}). To facilitate decoupling of the system, it is convenient to eliminate the radial dependence in one of the diagonal terms. This is achieved via the application of a unitary transformation
\begin{equation}
\mathfrak{U} = \exp\left(\frac{i}{2} \sigma_2 \rho \right)
= \begin{pmatrix}
\cos(\eta) & \sin(\eta) \\
-\sin(\eta) & \cos(\eta)
\end{pmatrix}.
\end{equation}
The implementation of this transformation leads to a reformulation of Eq.(\ref{hamprinc}) into the following system
\begin{equation}\label{mat2}
\resizebox{0.93\linewidth}{!}{$\displaystyle
\begin{pmatrix}
\cos(2\eta) + 2\alpha^2 V(r) - \epsilon &
-\sin(2\eta)+\frac{\alpha^2}{\sin(2\eta)} \left( \cos(2\eta)V(r) - U(r) \right) -\alpha \frac{d}{dr} \\\\
-\sin(2\eta)+\frac{\alpha^2}{\sin(2\eta)} \left( \cos(2\eta)V(r) - U(r) \right) +\alpha \frac{d}{dr} &
-\cos(2\eta) - \epsilon
\end{pmatrix}
\begin{pmatrix}
G_1(r) \\
G_2(r)
\end{pmatrix}
= 0,$}
\end{equation}
where the transformed spinor components are defined as
\begin{equation}
\begin{pmatrix}
G_1(r) \\
G_2(r)
\end{pmatrix}
= \mathfrak{U}
\begin{pmatrix}
F_1(r) \\
F_2(r)
\end{pmatrix}.
\end{equation}
The decoupling of the system can be achieved by selecting an effective vector potential of the form
\begin{equation}
A(r) = \frac{\alpha\cos(2\eta)}{\sin(2\eta)} \left[ \frac{V(r)}{\cos(2\eta)} - U(r) \right] - \frac{\lambda}{r} \left[ 1 + \alpha^2 U(r) \right],
\end{equation}
from which $G_{2}(r)$ can be obtained from Eq.~(\ref{mat2}) as
\begin{equation}\label{EG2}
G_{2}(r) = \frac{\alpha}{\cos(2\eta) + \epsilon} \left[ -\frac{\sin(2\eta)}{\alpha} + \frac{\alpha}{\sin(2\eta)} (\cos(2\eta) V(r) - U(r)) + \frac{d}{dr} \right] G_{1}(r).
\end{equation}
Given the relations \( V(r) = a z(r) \) and \( U(r) = b z(r) \), equation \eqref{mat2} leads to the following uncoupled differential equation for $G_{1}(r)$
\begin{equation}\label{Ecudpri}
\resizebox{0.93\linewidth}{!}{$\displaystyle
\left[ \frac{d^2}{dr^2} +  \frac{\alpha}{\sin(2\eta)}(a\cos(2\eta) - b)z'(r) - 2(b + \epsilon a)z(r) - \frac{\alpha^2}{\sin^2(2\eta)}(a\cos(2\eta) - b)^2 z^2(r) + \frac{\epsilon^2 - 1}{\alpha^2} \right] G_{1}(r) = 0.$}
\end{equation}
\section{Algebraic Treatment of the Hydrogen Atom in Curved space-time}
We consider the Coulomb potential \( V(r) = \frac{Z}{r} \), which implies \( z(r) = \frac{1}{r} \) and \( a = Z \). Under this assumption, Eq.~(41) becomes
\begin{equation}\label{secdifC}
\resizebox{0.93\linewidth}{!}{$\displaystyle
\left[ \frac{d^2}{dr^2} - \frac{\alpha}{\sin(2\eta)}(Z\cos(2\eta) - b)\left( \frac{\alpha}{\sin(2\eta)}(Z\cos(2\eta) - b) + 1 \right) \frac{1}{r^2} - 2(b + \epsilon Z) \frac{1}{r} + \frac{\epsilon^2 - 1}{\alpha^2} \right]G_{1}(r) = 0.$}
\end{equation}
To obtain the energy spectrum for this potential, we introduce the following set of operators
\begin{align}\label{B0}
\mathfrak{B}_0 &= \frac{1}{2}\left[-r\frac{d^2}{dr^2} + \frac{\frac{\alpha}{\sin(2\eta)}\left(Z\cos(2\eta) - b\right)\left(\frac{\alpha}{\sin(2\eta)}\left(Z\cos(2\eta) - b\right) + 1\right)}{r} + r\right],\\\label{B1}
\mathfrak{B}_1 &= \frac{1}{2}\left[-r\frac{d^2}{dr^2} + \frac{\frac{\alpha}{\sin(2\eta)}\left(Z\cos(2\eta) - b\right)\left(\frac{\alpha}{\sin(2\eta)}\left(Z\cos(2\eta) - b\right) + 1\right)}{r} - r\right],  \\\label{B2}
\mathfrak{B}_2 &= -i\rho\frac{d}{dr}.
\end{align}
These operators generate the Lie algebra \( su(1,1) \), similar to those introduced by Barut for central potentials~\cite{Bar1,Hecht1}. The corresponding Casimir operator is
\begin{equation}\label{Casah}
\mathfrak{C}^2_{HT} = \mathfrak{B}_0^2 - \mathfrak{B}_1^2 - \mathfrak{B}_2^2 = \frac{\alpha}{\sin(2\eta)}\left(Z\cos(2\eta) - b\right)\left(\frac{\alpha}{\sin(2\eta)}\left(Z\cos(2\eta) - b\right) + 1\right).
\end{equation}
By the representation theory of \( su(1,1) \), the eigenvalues of \( \mathfrak{C}^2 \) are given by
\begin{equation}
k_{HT}\left(k-1\right) = \frac{\alpha}{\sin(2\eta)}\left(Z\cos(2\eta) - b\right)\left(\frac{\alpha}{\sin(2\eta)}\left(Z\cos(2\eta) - b\right) +1\right),
\end{equation}
which implies that the Bargmann index \( k_{HT} \) is
\begin{equation}
k_{HT} = \frac{\alpha}{\sin(2\eta)}\left(Z\cos(2\eta) - b\right) + 1.
\end{equation}
In terms of the generators defined in Eqs.(\ref{B0}) and (\ref{B1}), Eq.(\ref{secdifC}) can be rewritten as
\begin{equation}
\left[-\left(\mathfrak{B}_0 + \mathfrak{B}_1\right) - 2\left(b + \epsilon Z\right) - \left(\mathfrak{B}_0 - \mathfrak{B}_1\right)\frac{\epsilon^2 - 1}{\alpha^2}\right] G_{1}(r)\equiv HG_{1}(r).
\end{equation}
To simplify the analysis, we perform a similarity transformation
\begin{align}
\widetilde{G}_{1}(r) &= e^{-i\varphi\mathfrak{B}_2} G_{1}(r), \\
\widetilde{H} &= e^{-i\varphi\mathfrak{B}_2} H e^{i\varphi\mathfrak{B}_2}.
\end{align}
Using the Baker-Campbell-Hausdorff formula, the transformation of the generators yields
\begin{align}
e^{-i\varphi\mathfrak{B}_2} \mathfrak{B}_0 e^{i\varphi\mathfrak{B}_2} &= \mathfrak{B}_0 \cosh(\varphi) + \mathfrak{B}_1 \sinh(\varphi), \\
e^{-i\varphi\mathfrak{B}_2} \mathfrak{B}_1 e^{i\varphi\mathfrak{B}_2} &= \mathfrak{B}_0 \sinh(\varphi) + \mathfrak{B}_1 \cosh(\varphi),
\end{align}
which leads to the general identity
\begin{equation}
e^{-i\varphi\mathfrak{B}_2}(\mathfrak{B}_0 \pm \mathfrak{B}_1)e^{i\varphi\mathfrak{B}_2} = e^{\pm\varphi}(\mathfrak{B}_0 \pm \mathfrak{B}_1).
\end{equation}
Therefore, the transformed Hamiltonian becomes
\begin{equation}
\widetilde{H} \widetilde{G}_1(r) = \left[
    \mathfrak{B}_0 \left( -e^{\varphi} - \frac{(\epsilon^2 - 1)e^{\varphi}}{\alpha^2} \right)
    + \mathfrak{B}_1 \left( -e^{\varphi} + \frac{(\epsilon^2 - 1)e^{\varphi}}{\alpha^2} \right)
    - 2(b + \epsilon Z)
\right] \widetilde{G}_1(r)=0.
\end{equation}
Choosing the scaling parameter as \(\varphi= \ln\left( \sqrt{\frac{\epsilon^2 - 1}{\alpha^2}} \right) \), the coefficient of \( \mathfrak{B}_1 \) vanishes, leading to the eigenvalue equation
\begin{equation}
\widetilde{H} \widetilde{G}_{l}(r) = \left[
    \mathfrak{B}_0 \left( -2 \sqrt{\frac{\epsilon^2-1}{\alpha^2}} \right) - 2(b + \epsilon Z)
\right] \widetilde{G}_{l}(r)=0,
\end{equation}
which simplifies to
\begin{equation}
\left[
    \mathfrak{B}_0 + \frac{b + \epsilon Z}{\sqrt{\frac{\epsilon^2 - 1}{\alpha^2}}}
\right] \widetilde{G}_{l}(r) = 0.
\end{equation}
Applying this operator identity to the $su(1,1)$ basis states (see Appendix, equation (\ref{k0n})) yields the condition
\begin{equation}
-\frac{b + \epsilon Z}{\sqrt{\frac{\epsilon^2 - 1}{\alpha^2}}} = k + n.
\end{equation}
From this, the energy spectrum for the hydrogen atom in curved space-time with a Coulomb potential can be derived
\begin{equation}
\epsilon_{nHT} = \frac{
    -\alpha^2 \mu z
    - \left(
        n + 1 + \frac{\alpha (Z \cos(2\eta) - b)}{\sin(2\eta)}
    \right)
    \sqrt{ \alpha^2 (\mu^2 - z^2) + \left(
        n + 1 + \frac{\alpha (Z \cos(2\eta) - b)}{\sin(2\theta)}
    \right)^2 }
}{
    \alpha^2 z^2 - \left(
        n + 1 + \frac{\alpha (Z \cos(2\eta) - b)}{\sin(2\eta)}
    \right)^2
}.
\end{equation}
The corresponding radial component, denoted \( G_{1}(r) \), is obtained from the second-order differential equation~\cite{Magh}
\begin{equation}\label{Soledif}
\frac{d^2G_{1}(r)}{dr^2}+\left(\frac{A_r}{r}+\frac{B_r}{r^2}+C_r\right)G_{1}(r)=0,
\end{equation}
whose solution is given by
\begin{equation}\label{Hass}
G_{1}(r)=r^{\frac{1}{2}+\sqrt{\frac{1}{4}-B_r}}e^{-\sqrt{-C_r}r}L_n^{2\sqrt{\frac{1}{4}-B_r}}\left(2\sqrt{-C_r}r\right),
\end{equation}
with
\begin{align}
A_r &= -2(b + \epsilon Z), \\
B_r &= \frac{\alpha}{S}(Z\cos(2\eta) - b)\left( \frac{\alpha}{\sin(2\eta)}(Z\cos(2\eta) - b) + 1 \right), \\
C_r &= \frac{1-\epsilon^2 }{\alpha^2}.
\end{align}
Thus, the normalized radial eigenfunction for \( G_{1}(r) \) is\cite{Oliv2,Magh}
\begin{equation}\label{G1H}
G_{1HT}(r) = r^{\gamma_{HT} + 1} e^{-\beta_{HT} r} L_n^{2\gamma_{HT} + 1}(2\beta_{HT} r),
\end{equation}
while the second spinor component \( G_2(r) \) is obtained via Eq.~(\ref{EG2}) and takes the form
\begin{equation}\label{G2H}
\begin{split}
G_{2HT}(r) &= r^{\gamma_{HT} + 1} e^{-\beta_{HT} r} \left[
    \left( -\frac{\sin(2\eta)}{\alpha} + \frac{2\gamma_{HT} + 1}{r} - \beta_{HT} \right) L_n^{2\gamma_{HT} + 1}(2\beta_{HT} r)
\right. \\
&\left.
    - 2\beta_{HT} r \, L_{n-1}^{2\gamma_{HT} + 2}(2\beta_{HT} r)
\right],
\end{split}
\end{equation}
where the parameters are defined as
\begin{equation}
\beta_{HT} = \sqrt{\frac{1 - \epsilon^2}{\alpha^2}},\quad
\gamma_{HT} = \frac{\alpha}{\sin(2\eta)}(Z\cos(2\eta) - b), \quad
\mathcal{N}_{HT} = \frac{(2\beta_{HT})^{\gamma_{HT} + 1/2}}{\sqrt{\mathcal{W}_1 + \alpha^2 b \beta_{HT} \, \mathcal{W}_2}},
\end{equation}
where $\mathcal{N}_{HT}$ is the normalization constant for the spinor wave functions $G_{1HT}(r)$ and $G_{2HT}(r)$, and where the explicit forms of $\mathcal{W}_1$ and $\mathcal{W}_2$ used are given in Eqs.~(\ref{W1AH}) and (\ref{W2AH}) of the Appendix.
\section{Algebraic Treatment of the Morse Oscillator in Curved space-time}
We now consider the Dirac-Morse oscillator, a key model in molecular physics, employing an algebraic method based on the symmetry of the Lie algebra \( su(1,1) \), in analogy with our previous analysis. The radial component satisfies the second-order differential equation
\begin{equation}\label{difMO}
\begin{split}
&\left[\frac{d^{2}}{dr^{2}}
- \frac{\alpha^{2}}{\sin^2(2\eta)}(a\cos(2\eta) + b)^{2}e^{-2\delta r}
\right. \\
&+ \left\{
    \frac{\alpha\delta}{\sin(2\eta)}(a\cos(2\eta) + b) + 2(\epsilon a - b)
  \right\}e^{-\delta r}
\left.
+ \frac{\epsilon^{2} - 1}{\alpha^{2}}
\right]G_{1}(r) = 0.
\end{split}
\end{equation}
Introducing the substitution \( \rho = e^{-\delta r} \), Eq.~\eqref{difMO} becomes
\begin{equation}\label{secondM}
\begin{split}
&\left[\frac{d^{2}}{d\rho^{2}} + \frac{1}{\rho}\frac{d}{d\rho}
- \frac{\alpha^{2}}{\sin^2(2\eta)\delta^2}(a\cos(2\eta) + b)^{2}
\right. \\
&+ \left\{
    \frac{\frac{\alpha\delta}{\sin(2\eta)}(a\cos(2\eta) + b) + 2(\epsilon a - b)}{\rho\delta^2}
  \right\}
\left.
- \frac{1-\epsilon^{2}}{\alpha^{2}\delta^2 z^2}
\right]G_{1}(\rho) = 0.
\end{split}
\end{equation}
To algebraically determine the energy spectrum, we define the generators of the \( su(1,1) \) algebra
\begin{align}
\mathfrak{L}_0 &= \frac{1}{2} \left(-\rho\frac{d^2}{d\rho^2}- \frac{d}{d\rho}+ \frac{1 - \epsilon^2}{\delta^2 \alpha^2 \rho}+ \rho\right), \\
\mathfrak{L}_1 &= \frac{1}{2} \left(-\rho\frac{d^2}{d\rho^2}- \frac{d}{d\rho}+ \frac{1 - \epsilon^2}{\delta^2 \alpha^2 \rho}- \rho\right), \\
\mathfrak{L}_2 &= -i \left(\rho\frac{d}{d\rho}+\frac{1}{2}\right).
\end{align}
The corresponding Casimir operator takes the form
\begin{equation}
\mathfrak{C}^2_{MT} = \mathfrak{L}_0^2 - \mathfrak{L}_1^2 - \mathfrak{L}_2^2 =\frac{1 - \epsilon^2}{\alpha^2 \delta^2}-\frac{1}{4},
\end{equation}
whose eigenvalues must satisfy
\begin{equation}
\frac{1 - \epsilon^2}{\alpha^2 \delta^2}-\frac{1}{4}=k_{MT}\left(k_{MT}-1\right).
\end{equation}
Thus, the Bargmann index $k$, which characterizes the irreducible unitary representation of the Lie algebra $su(1,1)$, is given by
\begin{equation}
k_{MT} = \frac{1}{2} + \sqrt{\frac{1 - \epsilon^2}{\alpha^2 \delta^2}}.
\end{equation}
By evaluating the action of $\mathfrak{L}_0$ on the standard basis states of the algebra, Eq.~(\ref{k0n}) of the appendix leads to the following quantization condition
\begin{equation}
\frac{ \alpha(a\cos(2\eta) + b)/(\delta\sin(2\eta)) + 2(\epsilon a - b)/\delta^2 }
{ 2 \sqrt{ \alpha^2(a\cos(2\eta) + b)^2 / (\sin^2(2\eta) \delta^2) } }
= \frac{1}{2} + \sqrt{ \frac{1 - \epsilon^2}{\alpha^2 \delta^2} } + n,
\end{equation}
from which the energy spectrum of the Dirac Morse oscillator is derived as
\begin{equation}\label{EMORS1}
\epsilon_{MT} = \frac{2 \Gamma a (\Gamma b + n \alpha \delta) \pm 2 \sqrt{ \Gamma^2 a^2 - \Gamma^2 b^2 - 2 n \alpha \delta \Gamma b - n^2 \alpha^2 \delta^2 + 1 }}
{2 (\Gamma^2 a^2 + 1)},
\end{equation}
with the parameter
\begin{equation}
\Gamma = \frac{\sin(2\eta)}{a \cos(2\eta) + b}.
\end{equation}
To construct the radial eigenfunctions, we apply the transformation $G_1(\rho) = \rho^{-1/2} F_1(\rho)$ to Eq.~\eqref{secondM}, which yields the differential equation
\begin{align}\label{EDMOR}
\left[\frac{d^{2}}{d\rho^{2}} - \frac{\alpha^{2}}{\sin^2(2\eta)\delta^{2}}(a\cos(2\eta)+ b)^{2}
+ \left\{ \frac{\alpha \delta (a\cos(2\eta) + b)/(\sin(2\eta)) + 2(\epsilon a - b)}{\delta^{2}} \right\} \frac{1}{\rho} \right. \notag \\
\left. + \left( \frac{1}{4} - \frac{1 - \epsilon^{2}}{\alpha^{2}\delta^{2}} \right) \frac{1}{\rho^{2}} \right] F_1(\rho) = 0.
\end{align}
The radial solution for the upper component, $F_{1MT}(\rho)$, is then given by~\cite{Oliv2,Magh}
\begin{equation}\label{F1M}
F_{1MT}(\rho) = \rho^{\beta_{MT}} \, e^{-\gamma_{MT} \rho}
\, L_n^{2\beta_{MT}}
\left(\frac{\alpha (a\cos(2\eta) + b)}{\sin(2\eta)\delta} \rho\right),
\end{equation}
while the lower spinor component, $F_{2MT}(\rho)$, obtained via Eq.~(\ref{EG2}), takes the explicit form
\begin{align}\label{F2M}
F_{2MT}(\rho) &=
\frac{\alpha \, \rho^{\beta_{MT}} e^{-\gamma_{MT} \rho}}{\cos(2\eta)+ \epsilon}
\Big[ \bigg( -\frac{\sin(2\eta)}{\alpha} - \delta \beta_{MT}
+ \left( \frac{\alpha (a\cos(2\eta) - b)}{\sin(2\eta)} + \delta \gamma_{MT} \right) \rho \bigg)
L_n^{(\lambda)}(\gamma_{MT} \rho) \notag \\
&\quad + \delta \gamma_{MT} \rho \, L_{n-1}^{(\lambda + 1)}(\gamma_{MT} \rho) \Big],
\end{align}
with
\begin{equation}
\beta_{MT} = \sqrt{\frac{1 - \epsilon^2}{\alpha^2 \delta^2}}, \quad
\gamma_{MT} = \frac{\alpha (a\cos(2\eta) + b)}{\sin(2\eta) \delta}, \quad
\lambda = 2\beta_{MT}.
\end{equation}
The explicit expression for the normalization constant associated with the radial wave functions $F_{1MT}(\rho)$ and $F_{2MT}(\rho)$ is provided in Eq.~(\ref{FNM}) of the Appendix.

\section{Algebraic Treatment of the Linear Radial Potential in Curved space-time}
We now examine the linear radial potential \( V(r) = ar \) in curved space-time using an algebraic approach. Unlike the \( su(1,1) \) generators previously employed following Gerry's method~\cite{Gerry}, here we derive the generators via the Schr\"odinger factorization procedure, which also yields a realization of the \( su(1,1) \) Lie algebra.
We begin by applying this potential to Eq.~(\ref{Ecudpri}) under the specific choice \( z(r) = r \), leading to the differential equation
\begin{equation}
\begin{split}
&\left[\frac{d^{2}}{dr^{2}}
- \frac{\alpha^{2}}{\sin^2(2\eta)}(a\cos(2\eta)-b)^{2} \left\{ r^{2} + \frac{\sin^2(2\eta)(b + \epsilon a)}{\alpha^{2}(b - a\cos(2\eta))}r \right\}
\right. \\
&\left.
- \frac{\alpha}{\sin(2\eta)}(b - a\cos(2\eta))
+ \frac{\epsilon^{2} - 1}{\alpha^{2}}
\right] G_{1}(r) = 0,
\end{split}
\end{equation}
with \( a, b > 0 \). Completing the square by introducing the change of variable
\begin{equation}
y = r + \frac{\sin^2(2\eta)(b + \epsilon a)}{\alpha^{2}(b - a\cos(2\eta))^{2}},
\end{equation}
the equation becomes
\begin{equation} \label{ecdifL}
\left[ \frac{d^{2}}{dy^{2}} - \left( \frac{\alpha^{2}}{\sin^2(2\eta)}(b - a\cos(2\eta))^{2} y^{2} - \mathcal{V} + \frac{1 - \epsilon^{2}}{\alpha^{2}} \right) \right] G_{1}(y) = 0,
\end{equation}
where
\begin{equation}
\mathcal{V} = \frac{\sin^2(2\eta)(b + \epsilon a)}{\alpha^{2}(b - a\cos(2\eta))^{2}} - \frac{\alpha(b - a\cos(2\eta))}{\sin(2\eta)}.
\end{equation}
To implement Schr\"odinger factorization, we propose the operator form
\begin{equation} \label{edsL2}
\left( y\frac{d}{dr} + \mathbb{A}y^{2} + \mathbb{B} \right) \left( -y \frac{d}{dr} + \mathbb{C}y^{2} + \mathbb{F} \right) G_{1}(y) = \mathbb{G} G_{1}(y),
\end{equation}
where \( \mathbb{A}, \mathbb{B}, \mathbb{C}, \mathbb{F}, \mathbb{G} \) are constants determined by matching Eq.~(\ref{edsL2}) to Eq.~(\ref{ecdifL}). This yields
\begin{equation}
\mathbb{A}= \mathbb{C} = \pm \beta_{LS}, \quad \mathbb{B} = -\frac{\mathcal{V} - \frac{1 - \epsilon^{2}}{\alpha^{2}}}{2(\pm \beta)} - \frac{3}{2},\quad \mathbb{F} = -\frac{\mathcal{V} - \frac{1 - \epsilon^{2}}{\alpha^{2}}}{2(\pm \beta)} - \frac{1}{2}, \quad \mathbb{G} = -\mathbb{B}(\mathbb{B}+1),
\end{equation}
with \( \beta_{LS} = \frac{\alpha(b - a\cos(2\eta))}{\sin(2\eta)} \). Accordingly, the factorized form of the differential equation becomes
\begin{equation}
(J_{\mp} \mp 1) J_{\pm} G_{1}(y) = -\left( \frac{\mathcal{V} - \frac{1 - \epsilon^{2}}{\alpha^{2}}}{2\beta} \pm \frac{3}{2} \right) \left( \frac{\mathcal{V} - \frac{1 - \epsilon^{2}}{\alpha^{2}}}{2\beta} \pm \frac{1}{2} \right),
\end{equation}
where the generators of the \( su(1,1) \) algebra are given by
\begin{align}
J_{+} &= \frac{1}{2} \left(- y \frac{d}{dy} + \beta_{LS} y^{2} - 2J_0 - \frac{1}{2} \right), \\
J_{-} &= \frac{1}{2} \left( y \frac{d}{dy} + \beta_{LS} y^{2} - 2J_0 +\frac{1}{2} \right), \\\label{J0}
J_0 &= \frac{1}{4\beta_{LS}} \left( -\frac{d^{2}}{dy^{2}} + \beta^{2}_{LS} y^{2} \right) = \frac{\mathcal{V} - \frac{1 - \epsilon^{2}}{\alpha^{2}}}{4\beta_{LS}}.
\end{align}
The Casimir operator $\mathfrak{C}^{2}_{LS}$ of the \( su(1,1) \) algebra is defined as
\begin{equation}
\mathfrak{C}^{2}_{LS} = J_{0}(J_{0} + 1) - J_{-} J_{+} = -\frac{3}{16} = k_{LS}\left(k_{LS} - 1\right),
\end{equation}
from which we identify the Bargmann index as \( k_{LS} = 1/4 \). Using this value together with the explicit expression for \( J_0 \) of equation (\ref{J0}) together with equation (\ref{k0n}), we derive the quantization condition
\begin{equation} \label{relacikT}
\frac{1}{4\beta_{LS}} \left( \frac{(b + \epsilon a)^2}{\beta^{2}_{LS}} - \beta_{LS} - \frac{1 - \epsilon^{2}}{\alpha^{2}} \right) = \frac{1}{4} + n,
\end{equation}
where $n$ denotes the radial quantum number.
Solving Eq.~\eqref{relacikT} for the relativistic energy parameter \( \epsilon \), we obtain
\begin{equation}
\epsilon_{LS} = \frac{ -\alpha^{2} ab + \sqrt{ \beta^{2} \left[ 2\alpha^{2} \beta (2n + 1)(\alpha^{2}a^{2} + \beta^{2}) + \alpha^{2}(a^{2} - b^{2}) + \beta^{2} \right] } }{ \alpha^{2}a^{2} + \beta^{2} }.
\end{equation}
The corresponding normalized eigenfunctions are expressed in terms of Hermite polynomials~\cite{GMS}. The upper spinor component reads
\begin{equation}
G_{1LS}(r) = \exp \left[ -\frac{\nu}{2} \left( r + \frac{b + \epsilon a}{\nu^{2}} \right)^{2} \right] H_{n} \left[ \sqrt{\nu} \left( r + \frac{b + \epsilon a}{\nu^{2}} \right) \right],
\end{equation}
while the lower spinor component is given by
\begin{align}
G_{2LS}(r) &= -\frac{\alpha}{C + \epsilon} \Bigg\{
\pi H_{n} \left[ \sqrt{\nu} \left( r + \frac{b + \epsilon a}{\nu^{2}} \right) \right]
+ \sqrt{\nu} H_{n+1} \left[ \sqrt{\nu} \left( r + \frac{b + \epsilon a}{\nu^{2}} \right) \right] \Bigg\} \notag \\
&\quad \times \exp \left[ -\frac{\nu}{2} \left( r + \frac{b + \epsilon a}{\nu^{2}} \right)^{2} \right],
\end{align}
where \( \nu = \frac{\alpha}{\sin(2\eta)}(b - a\cos(2\eta)) > 0 \) and
\begin{equation}
\pi = \frac{\sin(2\eta)}{\alpha} + \frac{(b + \epsilon a)(1 - 2k)}{k^{2}}.
\end{equation}
The normalization constant associated with these solutions has been explicitly derived in Ref.~\cite{GMS}.

\section{Hydrogen Atom in curved space time: An Algebraic Method via Schr\"odinger Factorization}
In this section, we study the hydrogen atom in curved space-time by employing the Schr\"odinger factorization method, similar to the procedure used in the previous section. We propose the factorized form
\begin{equation}\label{Fasc}
\left(r\frac{d}{dr} +\mathscr{A} r + \mathscr{B}\right)\left(-r\frac{d}{dr} + \mathscr{C}r + \mathscr{F}\right)G_{1}(r) = \mathscr{G}G_{1}(r),
\end{equation}
By expanding Eq.~\eqref{Fasc} and matching terms with Eq.~\eqref{secdifC}, we obtain the following coefficients
\begin{equation}
\begin{aligned}
\mathscr{A}_{HS} &= \mathscr{C}_{HS} = \pm\sqrt{\frac{1-\epsilon^2}{\alpha^2}}, \quad
\mathscr{B}_{HS} = \frac{b + \epsilon z}{\pm a} - 1, \quad
\mathscr{F}_{HS} = \frac{b + \epsilon z}{\pm a}, \\
\mathscr{G}_{HS} &= -\Theta(\Theta + 1) - \mathscr{B}_{HS}(\mathscr{B}_{HS} + 1).
\end{aligned}
\end{equation}
Thus, Eq.\eqref{secdifC} can be recast in factorized form
\begin{equation}
(R_{\mp} \mp 1)R_{\pm} = -\Theta(\Theta+1) - \frac{(b + \epsilon z)}{a} \left( \frac{(b + \epsilon z)}{a} \mp 1 \right),
\end{equation}
where \( \Theta = \frac{\alpha}{\sin(2\eta)}(Z\cos(2\eta) - b) \), and the \(su(1,1)\) operators are defined as
\begin{align}
R_+ &= -r \frac{d}{dr} + \sqrt{\frac{1 - \epsilon^2}{\alpha^2}} r - R_0, \\
R_- &= r \frac{d}{dr} + \sqrt{\frac{1 - \epsilon^2}{\alpha^2}} r - R_0, \\\label{Toah}
R_0 &= \frac{1}{2\sqrt{\frac{1 - \epsilon^2}{\alpha^2}}} \left[ -r \frac{d^2}{dr^2} + \frac{\Theta(\Theta + 1)}{r} - \frac{\epsilon^2 - 1}{\alpha^2} r \right] \notag \\
&= \frac{-(b + \epsilon Z)}{2\sqrt{\frac{1 - \epsilon^2}{\alpha^2}}}.
\end{align}
The quadratic Casimir operator is given by
\begin{equation}
\mathfrak{C}^2_{HS} = -R_+ R_- + R_0^2 - R_0 = \eta(\eta + 1) = k_{HS}\left(k_{HS} - 1\right),
\end{equation}
which implies the Bargmann index
\begin{equation}
k_{HS} = \eta + 1 = \frac{\alpha}{\sin(2\eta)}(Z\cos(2\eta) - b) + 1.
\end{equation}
From Eqs.~\eqref{Toah} and (\ref{k0n}) in the Appendix, we obtain the quantization condition
\begin{equation}
\frac{-(b + \epsilon Z)}{2\sqrt{\frac{1 - \epsilon^2}{\alpha^2}}} = n + \frac{\alpha}{\sin(2\eta)}(Z\cos(2\eta) - b) + 1,
\end{equation}
from which the relativistic energy spectrum for the hydrogen atom in curved space-time is derived
\begin{equation}
\epsilon_{nHS} = \frac{
    -\alpha^2 \mu z
    - \left(
        n + 1 + \frac{\alpha (Z \cos(2\eta) - b)}{\sin(2\eta)}
    \right)
    \sqrt{ \alpha^2 (\mu^2 - z^2) + \left(
        n + 1 + \frac{\alpha (Z \cos(2\eta) - b)}{\sin(2\eta)}
    \right)^2 }
}{
    \alpha^2 z^2 - \left(
        n + 1 + \frac{\alpha (Z \cos(2\eta) - b)}{\sin(2\eta)}
    \right)^2
}.
\end{equation}
\section{Dirac Morse oscillator in Curved space-time: An Algebraic Method via Schr\"odinger Factorization}
Following an approach analogous to that used in Sections~5 and 6, in particular Section~6, we begin by rewriting Eq.~(\ref{EDMOR}) as
\begin{align}\label{secondM2}
\left[\frac{d^{2}}{d\rho^{2}} - \frac{\alpha^{2}}{\sin^2(2\eta)\delta^{2}}(a\cos(2\eta)+ b)^{2}
+ \left\{ \frac{\alpha \delta (a\cos(2\eta) + b)/(\sin(2\eta)) + 2(\epsilon a - b)}{\delta^{2}} \right\} \frac{1}{\rho} \right. \notag \\
\left. + \left( \frac{1}{4} - \frac{1 - \epsilon^{2}}{\alpha^{2}\delta^{2}} \right) \frac{1}{\rho^{2}} \right] F_1(\rho) = 0.
\end{align}
Comparing Eqs.~(\ref{Fasc}) and (\ref{secondM2}), we identify the constants \( \mathscr{A} \), \( \mathscr{B} \), \( \mathscr{C} \), \( \mathscr{F} \), and \( \mathscr{G} \) as
\begin{align}
\mathscr{A}_{\mathrm{MS}} &= \mathscr{C}_{\mathrm{MS}} = \pm\sqrt{\frac{\alpha(a\cos(2\eta)+b)}{\sin(2\eta)\delta}}, \quad
\mathscr{B}_{\mathrm{MS}} = -\frac{\alpha(a\cos(2\eta)+b) + 2(\epsilon a - b)}{2a\delta^2} - 1, \notag \\
\mathscr{F}_{\mathrm{MS}} &= -\frac{\alpha(a\cos(2\eta)+b) + 2(\epsilon a - b)}{2a\delta^2}, \quad
\mathscr{G}_{\mathrm{MS}} = \frac{1}{4} - \frac{1 - \epsilon^2}{\alpha^2 \delta^2} - \mathscr{B}_{\mathrm{MS}}(\mathscr{B}_{\mathrm{MS}} + 1).
\end{align}
Thus, Eq.(\ref{secondM2}) can be expressed in factorized form
\begin{equation}
(K_{\mp} \mp 1)K_{\pm} = \frac{1}{4} - \frac{1 - \epsilon^2}{\alpha^2 \delta^2} - \Lambda(\Lambda \pm 1),
\end{equation}
where \( \Lambda = \frac{\frac{\alpha\delta}{\sin(2\eta)}(a\cos(2\eta) + b) + 2(\epsilon a - b)}{2a\delta^2} \). The operators are defined as
\begin{align}
K_+ &= -z \frac{d}{dz} + \frac{\alpha (a\cos(2\eta) + b)}{S \delta} z - K_0, \\
K_- &= z \frac{d}{dz} + \frac{\alpha (a\cos(2\eta) + b)}{\sin(2\eta) \delta} z - K_0, \\\label{KM2}
K_0 &= \frac{1}{2} \left(-z \frac{d^2}{dz^2} + \frac{\alpha^2 (a\cos(2\eta) + b)^2}{\sin^2(2\eta)\delta^2} z - \left( \frac{1}{4} - \frac{1 - \epsilon^2}{\alpha^2 \delta^2} \right) \frac{1}{z} \right)\\ &= \frac{\frac{\alpha\delta}{\sin(2\eta)}(a\cos(2\eta) + b) + 2(\epsilon a - b)}{2a\delta^2}.
\end{align}
The quadratic Casimir operator \( \mathfrak{C}^2_{\mathrm{MS}} \) satisfies the eigenvalue equation
\begin{equation}\label{CasMo2}
\mathfrak{C}^2_{MS} = -K_+K_- + K_0^2 - K_0 = \frac{1 - \epsilon^2}{\alpha^2 \delta^2} - \frac{1}{4} = k_{MS}\left(k_{MS}-1\right),
\end{equation}
where the Bargmann index is given by
\begin{equation}\label{Bar2}
k_{MS} = \frac{1}{2} + \sqrt{\frac{1 - \epsilon^2}{\alpha^2 \delta^2}}.
\end{equation}
By applying Eqs.~(\ref{KM2}) to (\ref{Bar2}) together with Eq.~(\ref{k0n}) of the Appendix, the same energy spectrum previously obtained in Eq.~(\ref{EMORS1}) of Section~4 is recovered
\begin{equation}
\epsilon_{MS} = \frac{2 \Gamma a (\Gamma b + n \alpha \delta) \pm 2 \sqrt{\Gamma^2 a^2 - \Gamma^2 b^2 - 2 n \alpha \delta \Gamma b - n^2 \alpha^2 \delta^2 + 1}}{2 (\Gamma^2 a^2 + 1)}.
\end{equation}
\section{Concluding Remarks}
In this work, we have investigated the Dirac equation in static curved space-time, focusing on obtaining exact solutions through two distinct algebraic approaches.

The first method involves the introduction of a set of operators that close the $su(1,1)$ algebra. These operators are of the form originally introduced by Barut\cite{Bar1} in the context of generalized central potentials. Utilizing this framework, we applied the theory of unitary representations to derive both the energy spectrum and eigenfunctions for two representative cases: the hydrogen atom and the Dirac-Morse oscillator. This algebraic technique effectively yields exact solutions, demonstrating the utility of the $su(1,1)$ algebra in relativistic quantum systems.

The second approach is based on the Schr\"odinger factorization method. To employ this method, we first decouple the radial equations of motion, which allows the construction of three operators that also satisfy the $su(1,1)$ Lie algebra. With this structure, the theory of unitary irreducible representations again facilitates the derivation of exact energy spectra and the general form of relativistic spinor wave functions.

Unlike the first approach, which is restricted to the hydrogen atom and the Dirac-Morse oscillator, the Schr\"odinger factorization method exhibits broader applicability. It extends to three representative potentials: the hydrogen atom, the Dirac-Morse oscillator, and a linear radial potential. This generality enhances its relevance for solving a wider class of problems in relativistic quantum mechanics.

Finally, we emphasize that our results are consistent with those reported in the original reference on which this study is based\cite{GMS}

\appendix
\renewcommand{\theequation}{A.\arabic{equation}}
\setcounter{equation}{0}
\section{Appendix: $su(1,1)$ Algebra}
Within the framework of Lie algebra theory, the $su(1,1)$ algebra is generated by three operators, $\mathcal{J}_{\pm}$ and $\mathcal{J}_0$, which satisfy the following commutation relations~\cite{Vou}
\begin{equation}
[\mathcal{J}_0, \mathcal{J}_{\pm}] = \pm \mathcal{J}_{\pm}, \quad [\mathcal{J}_-, \mathcal{J}_+] = 2 \mathcal{J}_0. \label{comm}
\end{equation}
The action of these generators on the Fock space basis $\{|k, n\rangle,\; n = 0, 1, 2, \dots\}$ is given by
\begin{align}
\mathcal{J}_{+} |k, n\rangle &= \sqrt{(n+1)(2k+n)}\,|k, n+1\rangle, \label{k+n} \\
\mathcal{J}_{-} |k, n\rangle &= \sqrt{n(2k+n-1)}\,|k, n-1\rangle, \label{k-n} \\
\mathcal{J}_{0} |k, n\rangle &= (k + n)\,|k, n\rangle, \label{k0n} \\
\mathcal{C}^2 |k, n\rangle &= k(k - 1)\,|k, n\rangle. \label{Casape}
\end{align}

\section{Hydrogen atom}
\renewcommand{\theequation}{\thesection.\arabic{equation}}
\setcounter{equation}{0}
To compute the normalization constant for the hydrogen atom in curved space-time, we require that the spinor be normalized according to the following condition:
\begin{equation}\label{NAH}
\mathcal{N}^2 \int_0^\infty \left(|G_1(r)|^2 + |G_2(r)|^2\right)\left(1 + \alpha^2 U(r)\right)\,dr = 1,
\end{equation}
where \( U(r) = \frac{b}{r} \), and \( G_1(r) \) and \( G_2(r) \) are given by Eqs.(\ref{G1H}) and (\ref{F1M}), respectively. Substituting these expressions into Eq.(\ref{NAH}), and employing integral identities involving Laguerre polynomials specifically, Eqs.~17 and 19 from Section 2.19.14 of Ref.\cite{Prud}, and the orthogonality properties from Section 7.414 of Ref.\cite{Gran} we obtain the normalization constant:
\begin{equation}
\mathcal{N} = \frac{(2\beta_{HT})^{\gamma + 1/2}}{\sqrt{\mathcal{W}_1 + \alpha^2 b \beta_{HT} \, \mathcal{W}_2}},
\end{equation}
where the auxiliary quantities \(\mathcal{W}_1\) and \(\mathcal{W}_2\) are given by
\begin{equation}\label{W1AH}
\mathcal{W}_1 = \frac{(2\gamma + 2)_n}{n!} \left[ 1 + \left(\frac{\sin(2\theta)}{\alpha \beta_{HT}} - \frac{2\gamma + 1}{2}\right)^2 \right] + \frac{4\beta^2_{HT} (2\gamma + 3)_n}{(2\gamma + 1) n!},
\end{equation}
\begin{equation}\label{W2AH}
\mathcal{W}_2 = \frac{(2\gamma + 1)_n}{n!} \left[ 2 + \frac{(2\gamma + 1)^2}{2} \left(1 - \frac{\sin(2\eta)}{\alpha \beta_{HT} (2\gamma + 1)}\right)^2 \right] + \frac{4\beta^2_{HT} (2\gamma + 2)_n}{(2\gamma + 1) n!}.
\end{equation}
Here, \((x)_n\) denotes the Pochhammer symbol, which is defined in terms of the gamma function as
\begin{equation}
(x)_n = \frac{\Gamma(x + n)}{\Gamma(x)}.
\end{equation}
\section{Dirac Morse}
\renewcommand{\theequation}{\thesection.\arabic{equation}}
\setcounter{equation}{0}
In order to calculate the normalization constant for this system, we impose the following condition
\begin{equation}\label{NMO}
\mathcal{N}^2 \int_0^1 \left( |F_1(\rho)|^2 + |F_2(\rho)|^2 \right) \left(1 + \alpha^2 b \rho\right) \frac{1}{\delta \rho} \, d\rho = 1,
\end{equation}
where the substitution \(\rho = e^{-\delta r}\) has been applied. To obtain the normalization constant \(\mathcal{N}\), we proceed as follows: first, we substitute the radial wave functions \(F_1(\rho)\) and \(F_2(\rho)\), given in Eqs.~(\ref{G1H}) and (\ref{G2H}), into Eq.~(\ref{NMO}), considering the potential \(U(r) = b\rho\).
We then apply key integral identities involving Laguerre polynomials from Prudnikov's formulas (Eqs.~17 and 19 in Section 2.19.14 of Ref.~\cite{Prud}) along with the orthogonality relations from Section 7.414 of Ref.~\cite{Gran}. This procedure yields the final expression for \(\mathcal{N}\) \cite{GMS}
\begin{equation}\label{FNM}
\begin{aligned}
\mathcal{N} = \Bigg[ &\mathcal{K}_1
\Bigl( I_{n,n}(2\nu, 2\nu, 2\nu) + \alpha^2 b \, I_{n,n}(2\nu-1, 2\nu, 2\nu) \Bigr) \\
& - \mathcal{K}_2(n+2\nu)
\Bigl( I_{n,n-1}(2\nu, 2\nu, 2\nu) + \alpha^2 b \, I_{n,n-1}(2\nu-1, 2\nu, 2\nu) \Bigr) \\
& + \mathcal{K}_3(n+2\nu)^2
\Bigl( I_{n-1,n-1}(2\nu, 2\nu, 2\nu) + \alpha^2 b \, I_{n-1,n-1}(2\nu-1, 2\nu, 2\nu) \Bigr) \Bigg]^{-1/2},
\end{aligned}
\end{equation}
where the coefficients are given by
\begin{equation}
\mathcal{K}_1 = 1 + \frac{4\alpha^2 a^2}{\delta^2 \mu^2}, \qquad
\mathcal{K}_2 = -\frac{4\alpha^2 a}{\mu(\cos(2\eta) + \epsilon)}, \qquad
\mathcal{K}_3 = \frac{\alpha^2 \delta^2}{(\cos(2\eta) + \epsilon)^2},
\end{equation}
and the integrals \(I_{m,n}(\alpha, \beta, \gamma)\) are defined as
\begin{equation}
I_{m,n}(\alpha, \beta, \gamma) = \int_{0}^{1} \rho^{\alpha - 1} (-\ln \rho)^{\lambda} e^{-\mu \rho} L_m^{\beta}(\mu \rho) L_n^{\gamma}(\mu \rho) \, d\rho,
\end{equation}
with the parameters
\begin{equation}
\rho = e^{-\delta r}, \qquad \lambda = \frac{\alpha}{\delta} - 1, \qquad \mu = \frac{\alpha(a\cos(2\eta) + b)}{\sin(2\eta) \delta}.
\end{equation}
\textbf{\Large{Acknowledgments}}\\
This work was partially supported by SNII-M\'exico, CONAHCYT,
COFAA-IPN, EDI-IPN, SIP-IPN project numbers 20242284 and 20241764.

\end{document}